\documentclass[aps,amsfonts,amsmath,prd,nofootinbib,tightenlines,superscriptaddress,10pt]{revtex4}

\usepackage{epsfig}
\usepackage{bm}
\usepackage{bbm}

\newcommand{\kappabar}{\bar \kappa}
\newcommand{\wavenumbersq}{\kappa^2 + k_z^2}
\newcommand{\wavenumber}{\sqrt{\wavenumbersq}}
\newcommand{\wavevars}{\kappa,k_z}
\newcommand{\dsqdzsq}{\frac{\partial^2}{\partial z^2}}

\begin{document}
\title{Renormalized Quantum Stress-Energy Tensor of a 
Nonzero Radius Cosmic String}

%\pacs{03.65.Nk 11.10.Gh 11.27.+d 11.55.Hx}

\author{Noah Graham}
\email{ngraham@middlebury.edu}
\affiliation{Department of Physics, Middlebury College,
Middlebury, Vermont 05753, USA}

\begin{abstract}
We calculate the effects of quantum fluctuations of a
scalar field in the ``ballpoint pen'' cosmic string geometry.
Using the approach to renormalization established previously for the
energy density in two space dimensions, we extend those calculations
to $3+1$ dimensions, nonzero scalar mass, and the full stress-energy
tensor, including its contribution to the null energy condition for
radial geodesics.  The calculation demonstrates in detail the process of
renormalization in curved spacetime, including the effects of the
conformal anomaly.  This model provides one of the few examples
where quantum effects in curved spacetime can be explicitly
calculated.
\end{abstract}

\maketitle

\section{Introduction}

Curved space backgrounds present unique challenges for
renormalization in quantum field theory.  In topologically nontrivial
backgrounds, such calculations must compare fluctuations between
globally distinct sectors, while consistently imposing precise
renormalization conditions.  In a $3+1$ dimensional model, one must go
beyond renormalization of the Einstein-Hilbert term $\displaystyle
\frac{{\cal R}}{8 \pi G}$, written in terms of the Ricci scalar ${\cal
  R}$ and gravitational constant $G$, to include higher-order
counterterms proportional to ${\cal R}^2$ and $\Box {\cal R}$ in the
effective action.  Although the effects of these terms are too small 
for their coefficients to be measured, they play an important role in
fundamental features of quantum field theory in curved spacetime,
including the conformal anomaly \cite{BrlDv}.

Cosmic string geometries
\cite{PhysRevD.34.1918,PhysRevD.35.536,PhysRevD.35.3779,
PhysRevD.31.3288,1985ApJ...288..422G,PhysRevD.42.2669}
can provide a valuable theoretical laboratory in which to study these
effects.  With curvature only in the spatial $r$-$\theta$ plane
perpendicular to the string axis, they provide a geometry in which it
is tractable to compute the scattering data required to analyze
quantum fluctuations.  At the same time, they demonstrate topological
effects through a deficit angle that persists at large distances.
Such calculations were first carried out in the ``point string''
model, where one approximates the string to have infinitesimal
thickness, with a corresponding divergence in the curvature yielding a
finite, nonzero integral over the string cross section
\cite{PhysRevD.34.1918,PhysRevD.35.536,PhysRevD.35.3779}.  To remove
this unphysical idealization, the curvature can be spread 
over nonzero thickness by introducing a string radius $r_0$.  In the
``flowerpot'' model, the curvature is localized to a cylindrical surface
at $r=r_0$, while the ``ballpoint pen'' model has constant nonzero
curvature over a solid cylinder from $r=0$ to $r=r_0$
\cite{PhysRevD.31.3288,1985ApJ...288..422G,PhysRevD.42.2669}.  
While both can be treated by the methods described here, the ballpoint
pen is of greater interest because it is both fully nonsingular and
contains extended regions (rather than isolated singularities) at
which the curvature is nonzero, leading to contributions from
renormalization counterterms. 

This paper extends techniques previously applied to the quantum energy
density of a massless scalar field in the background of a 
cosmic string in $2+1$ dimensions \cite{PhysRevD.110.105009} to
compute the full renormalized stress-energy tensor in a
$3+1$-dimensional model.  In two space dimensions the geometry is
conformally flat, and renormalization can be expressed entirely in
terms of the $1+1$-dimensional trace anomaly
\cite{PhysRevD.13.2720,1977RSPSA.354...59D,PhysRevD.15.2088},
which is linear in ${\cal R}$.  Extending to $3+1$ dimensions then
requires the introduction of quadratic counterterms.  By first
generalizing to a massive particle, this calculation shows how the
conformal anomaly emerges from the second-order renormalization
counterterm through a quantum contribution to the trace of the
stress-energy tensor that persists even as the scalar mass goes to zero.

After introducing the field theory model in Sec.\ \ref{sec:model} and
the associated scattering theory in Sec.\ \ref{sec:scattering}, 
in Sec.\ \ref{sec:renorm} we establish the key tools for the
calculation, which give the connection developed in 
Ref.\ \cite{OLUM2003175} between quantum expectation values and
analytic scattering data, as carried out in dimensional
regularization for a background that is spherically
symmetric in $m$ dimensions and translationally invariant in $n$
dimensions.  The heat kernel coefficients used to
define renormalization in curved spacetime then emerge from the
leading terms in the Born approximation.  To make the calculation
numerically tractable, we add and subtract the result for the point
string as an intermediate result in Sec.\ \ref{sec:point}, and
illustrate particular subtleties of this process for the calculation of the
difference between the radial and angular pressures in Sec.\
\ref{sec:diff}.  We show numerical results in Sec.\ \ref{sec:results},
and summarize the calculation in Sec.\ \ref{sec:conclusions}.

\section{Model and Green's Function}
\label{sec:model}

We consider a free scalar field $\phi$ of mass $\mu$ in $3+1$ spacetime
dimensions, for which the action functional is
\begin{equation}
S = -\frac{1}{2}\int d^d x \sqrt{-g} \left(
\nabla_\alpha \phi \nabla^\alpha \phi + \xi {\cal R} \phi^2
+ \mu^2 \phi^2\right) \,,
\end{equation}
including coupling $\xi$ to the Ricci curvature scalar ${\cal R}$.  Of
particular interest is the case of conformal coupling, $\displaystyle
\xi = \frac{1}{6}$.  The equation of motion is
\begin{equation}
-\nabla_\alpha\nabla^\alpha\phi + \xi {\cal R} \phi + \mu^2 \phi= 0
\end{equation}
with metric signature $(-+++)$.  The stress-energy tensor is given by
\cite{PhysRevD.54.6233,PhysRevD.72.065013,fliss2023}
\begin{equation}
T_{\alpha\beta} = \nabla_\alpha \phi \nabla_\beta \phi
-g_{\alpha \beta} \frac{1}{2} \left(\nabla_\gamma \phi \nabla^\gamma \phi
+\mu^2 \phi^2 \right)
+ \xi \phi^2 \left( R_{\alpha\beta} -\frac{1}{2} g_{\alpha \beta} {\cal R}
\right)
+ \xi \left(g_{\alpha \beta} \nabla_\gamma \nabla^\gamma
- \nabla_\alpha \nabla_\beta\right)\phi^2\,,
\label{eqn:stressenergy}
\end{equation}
as obtained by varying the action with respect to the metric.  Note
that the curvature coupling contributes to the stress-energy tensor even
in regions where ${\cal R}=0$, although it does so by a total derivative.

We consider the spacetime metric
\cite{PhysRevD.31.3288,1985ApJ...288..422G,PhysRevD.42.2669}
\begin{equation}
ds^2 = -dt^2 + p(r)^2 dr^2 + r^2 d\theta^2 + dz^2
\end{equation}
with a deficit angle $2\theta_0$, meaning that the
range of angular coordinate is $0\ldots 2(\pi-\theta_0)$, and we
define $\displaystyle \sigma = \frac{\pi}{\pi-\theta_0}$.  To
implement the deficit angle without a singularity at the origin, we
introduce a profile function $p(r)$ that ranges from 
$\displaystyle \frac{1}{\sigma}$ at the origin
to $1$ at the string radius $r_0$.  The
nonzero Christoffel symbols in this geometry are
\begin{equation}
\Gamma_{r r}^r =  \frac{p'(r)}{p(r)} \qquad
\Gamma_{\theta \theta}^r = -\frac{r}{p(r)^2} \qquad
\Gamma_{\theta r}^\theta = \Gamma_{r \theta}^\theta = \frac{1}{r}\,,
\end{equation}
and because the geometry only has curvature in two dimensions, all the
nonzero components of the Riemann and Ricci tensors
\begin{equation}
R_{\theta r\theta}^r = -R_{\theta\theta r}^r = 
R_{\theta\theta} = g_{\theta\theta} \frac{{\cal R}}{2}
\qquad
R_{r\theta r}^\theta = -R_{rr\theta}^\theta
=  R_{rr} = g_{rr} \frac{{\cal R}}{2}
\label{eqn:curvature}
\end{equation}
can be expressed in terms of the curvature scalar $\displaystyle
{\cal R}=\frac{2}{r} \frac{p'(r)}{p(r)^3}$.  It obeys the
Gauss-Bonnet theorem in the $r$-$\theta$ plane,
\begin{equation}
\int_0^{r_0} p(r) dr \int_0^{2\pi/\sigma} r d\theta \, {\cal R}
= \frac{4\pi}{\sigma} \int_0^{r_0} \frac{p'(r)}{p(r)^2} dr =
\left(-\frac{4\pi}{\sigma} \frac{1}{p(r)} \right|_{r=0}^{r=r_0}
= 4\pi \left(1-\frac{1}{\sigma}\right) = 4 \theta_0 \,,
\end{equation}
for any $p(r)$ obeying the boundary conditions given above.

Acting on any scalar $\chi$, the covariant derivatives simply become
ordinary derivatives, while for second derivatives we have nontrivial
contributions from the Christoffel symbols given above,
\begin{equation}
\nabla_\theta \nabla_\theta \chi = \partial_\theta^2 \chi -
\Gamma_{\theta\theta}^r \partial_r \chi \qquad
\nabla_r \nabla_r \chi = \partial_r^2 \chi -
\Gamma_{rr}^r \partial_r \chi \qquad
\nabla_r \nabla_\theta \chi =
\nabla_\theta \nabla_r \chi = \partial_\theta \partial_r \chi -
\Gamma_{r\theta}^\theta \partial_\theta \chi \,,
\end{equation}
and, as a result, covariant derivatives with respect to $\theta$
can be nonzero even if $\chi$ is rotationally invariant.  In
particular, we have
\begin{equation}
(g^{\theta \theta} \nabla_\theta \nabla_\theta + g^{rr} \nabla_r
\nabla_r) \chi = \frac{1}{r^2} \left(\frac{\partial^2 \chi}
{\partial \theta^2} + {\cal D}_r^2 \right)\chi,
\end{equation}
where $\displaystyle {\cal D}_r = \frac{r}{p(r)} \frac{\partial}{\partial r}$
is the radial derivative.  The equation of motion for $\phi$ then becomes
\begin{equation}
\left(\frac{\partial^2}{\partial t^2} - \frac{1}{r^2} {\cal D}_r^2
 - \frac{1}{r^2} \frac{\partial^2}{\partial \theta^2} 
- \frac{d^2}{dz^2} + \mu^2 + \xi {\cal R}\right) \phi = 0\,,
\label{eqn:eomphi}
\end{equation}
which in combination with 
$\displaystyle {\cal D}_r^2 (\phi^2) = 
2 ({\cal D}_r \phi)^2 + 2 \phi {\cal D}_r^2 \phi$
gives the relation between expectation values
\begin{equation}
\left\langle \frac{1}{4 r^2} {\cal D}_r^2 (\phi^2) \right\rangle =
\left\langle \frac{1}{2r^2}({\cal D}_r \phi)^2 
-\frac{1}{2} (\partial_t \phi)^2 + \frac{1}{2r^2}
(\partial_\theta \phi)^2 + \frac{1}{2}(\partial_z \phi)^2
+ \frac{\mu^2}{2} \phi^2 + \frac{\xi}{2} {\cal R}\phi^2 
\right\rangle - {\cal A}\,,
\label{eqn:anomalousrelation}
\end{equation}
where ${\cal A}$ is an anomalous contribution discussed below
and we have used $\langle (\partial_t \phi)^2 \rangle = 
- \langle \phi (\partial_t^2 \phi) \rangle $, and similarly for 
the $\theta$ and $z$ derivatives.  We therefore obtain the
stress-energy tensor
\begin{eqnarray}
\langle T_t^t \rangle &=& -\langle {\cal H} \rangle =
-\left\langle \frac{1}{2} (\partial_t \phi)^2 
+\frac{1}{2r^2}({\cal D}_r \phi)^2
+ \frac{1}{2r^2}(\partial_\theta \phi)^2 + 
\frac{1}{2}(\partial_z \phi)^2 + \frac{\mu^2}{2} \phi^2
+ \frac{\xi}{2} {\cal R}\phi^2 - \frac{\xi}{r^2} {\cal D}_r^2
(\phi^2) \right\rangle \cr
&=& -\left\langle (\partial_t \phi)^2 + \left(\frac{1}{4}- \xi \right) 
\frac{1}{r^2} {\cal D}_r^2 (\phi^2) \right\rangle + {\cal A} \cr
\langle T_r^r \rangle &=&  \left\langle
\frac{1}{2} (\partial_t \phi)^2 + \frac{1}{2r^2}({\cal D}_r \phi)^2
- \frac{1}{2r^2}(\partial_\theta \phi)^2 -\frac{1}{2}(\partial_z \phi)^2
- \frac{\mu^2}{2} \phi^2
+ \frac{\xi}{r^2} \frac{1}{p(r)} {\cal D}_r (\phi^2) \right\rangle \cr
&=& \left\langle \frac{1}{r^2}({\cal D}_r \phi)^2 - 
\frac{1}{4r^2} {\cal D}_r^2 (\phi^2)
+ \frac{\xi}{r^2} \frac{1}{p(r)} {\cal D}_r (\phi^2) 
+ \frac{\xi}{2} {\cal R}\phi^2 \right\rangle + {\cal A} \cr
\langle T_\theta^\theta \rangle &=& \left\langle
\frac{1}{2} (\partial_t \phi)^2 - \frac{1}{2r^2}({\cal D}_r \phi)^2
+ \frac{1}{2r^2}(\partial_\theta \phi)^2 -\frac{1}{2}(\partial_z \phi)^2
- \frac{\mu^2}{2} \phi^2
+\frac{\xi}{r^2} \left(
{\cal D}_r^2 (\phi^2) - \frac{1}{p(r)} {\cal D}_r (\phi^2)\right)
 \right\rangle \cr
&=& \left\langle \frac{1}{r^2}
(\partial_\theta \phi)^2 - \left(\frac{1}{4} - \xi \right)
\frac{1}{r^2} {\cal D}_r^2 (\phi^2) 
-\frac{\xi}{r^2} \frac{1}{p(r)}{\cal D}_r (\phi^2) 
+ \frac{\xi}{2} {\cal R}\phi^2 \right\rangle + {\cal A} \cr
\langle T_z^z \rangle &=& \left\langle
\frac{1}{2} (\partial_t \phi)^2 - \frac{1}{2r^2}({\cal D}_r \phi)^2
- \frac{1}{2r^2}(\partial_\theta \phi)^2 + 
\frac{1}{2}(\partial_z \phi)^2 - \frac{\mu^2}{2} \phi^2
- \frac{\xi}{2} {\cal R}\phi^2 + \frac{\xi}{r^2} {\cal D}_r^2
(\phi^2)  \right\rangle \cr
&=& \left\langle (\partial_z \phi)^2 - \left(\frac{1}{4}- \xi \right) 
\frac{1}{r^2} {\cal D}_r^2 (\phi^2) \right\rangle + {\cal A}\,
\end{eqnarray}
where we have used mixed indices to simplify metric factors and
$\langle {\cal H} \rangle$ denotes the energy density.  By Lorentz
symmetry in the $z$ direction, we have $\langle (\partial_t \phi)^2
\rangle =  -\langle (\partial_z \phi)^2 \rangle$ and 
$\langle T_t^t \rangle = \langle T_z^z \rangle$.

We define the Green's function $G_\sigma(\bm{r},\bm{r}',\wavevars)$
for imaginary wave number $k=i\kappa$ and transverse momentum $k_z$,
which obeys
\begin{equation}
\left(-\frac{1}{r^2} {\cal D}_r^2
- \frac{1}{r^2} \frac{\partial^2}{\partial \theta^2}
-\dsqdzsq + \xi {\cal R} + \wavenumbersq \right)
G_\sigma(\bm{r},\bm{r}',\wavevars)  = \frac{1}{r p(r)} 
\delta(r-r') \delta(\theta-\theta') e^{ik_z(z-z')} \,,
\end{equation}
and we consider the ``ballpoint pen'' profile function 
\begin{equation}
p(r) = 
\begin{cases}
\left[\sigma^2-\frac{r^2}{r_0^2} (\sigma^2-1)\right]^{-1/2}&r<r_0\\
1 &r>r_0\end{cases}\,,
\end{equation}
where $r_0$ is the string radius, which has 
constant curvature
$\displaystyle {\cal R}=\frac{2 (\sigma^2-1)}{r_0^2}$
inside and zero curvature outside.
The ``flowerpot'' model, with zero curvature everywhere except for a
$\delta$-function contribution at the string radius, can also be
calculated straightforwardly using the same techniques, but that case
does not fully demonstrate the renormalization process because the
curvature is zero for all $r\neq r_0$, and as a result after subtracting
the free contribution the result is already finite, with all other
counterterms vanishing.

\section{Scattering Wave Functions}
\label{sec:scattering}

We write the Green's function in the scattering form
\begin{equation}
G_\sigma(\bm{r},\bm{r}',\wavevars) = \frac{\sigma}{\pi}
\sum_{\ell=0}^\infty{}'
\psi_{\wavevars,\ell}^{\rm reg} (r_<) \psi_{\wavevars,\ell}^{\rm out} (r_>)
\cos \left[\sigma \ell (\theta-\theta')\right] e^{i k_z (z-z')}\,,
\end{equation}
where the prime on the sum indicates that the $\ell=0$ term is counted
with a weight of one-half, arising because we have written the sum
over nonnegative $\ell$ only.  The radial wavefunctions obey the equation
\begin{equation}
\left(-\frac{1}{r^2} {\cal D}_r^2 + 
\frac{\ell^2\sigma^2}{r^2} + \xi {\cal R} + \wavenumbersq
\right)\psi_{\wavevars,\ell}(r) = 0 \,,
\label{eqn:eom}
\end{equation}
where the regular solution is
defined to be well-behaved at $r=0$, while the outgoing solution
obeys outgoing wave boundary conditions for $r\to \infty$, normalized
to unit amplitude.  Here $r_<$ ($r_>$) is the smaller (larger) axial
radius of $\bm{r}$ and $\bm{r}'$.  The regular functions are normalized
so that they obey the Wronskian relation
\begin{equation}
\frac{d}{dr} \left(\psi_{\wavevars,\ell}^{\rm reg}(r)\right)
\psi_{\wavevars,\ell}^{\rm out}(r)
- \psi_{\wavevars,\ell}^{\rm reg}(r) \frac{d}{dr} 
\left(\psi_{\wavevars,\ell}^{\rm out}(r)\right)
= \frac{p(r)}{r} \,,
\label{eqn:wronk}
\end{equation}
which provides the appropriate jump condition for the Green's function.

Next we construct the scattering wavefunctions in the string
background.  As shown in Ref.\ \cite{PhysRevD.42.2669},
we can write the full regular and outgoing solutions
$\psi_{\wavevars,\ell}(r)$ in terms of Legendre and Bessel functions as
\begin{equation}
\psi_{\wavevars,\ell}(r) =
\begin{array}{|c|c|c|}
\hline
& r<r_0 & r>r_0\cr
\hline
\hbox{regular} & A_{\wavevars,\ell}
P_{\nu(\wavevars)}^\ell\left(\tfrac{1}{\sigma  p(r)}\right) & 
I_{\sigma \ell}(\wavenumber r) + B_{\wavevars,\ell}
K_{\sigma \ell}(\wavenumber r)
\cr
\hline
\hbox{outgoing} &  
C_{\wavevars,\ell}
P_{\nu(\wavevars)}^\ell\left(\tfrac{1}{\sigma p(r)}\right) +
D_{\wavevars,\ell}
Q_{\nu(\wavevars)}^\ell\left(\tfrac{1}{\sigma p(r)}\right)
& K_{\sigma \ell}(\wavenumber r) \cr
\hline
\end{array}
\end{equation}
with
\begin{equation}
\nu(\wavevars) = -\frac{1}{2} + \frac{1}{2}\sqrt{(1-8\xi) - 
\frac{4(\wavenumbersq) r_0^2}{\sigma^2-1}} \,.
\end{equation}
Matching boundary conditions at $r=r_0$, we obtain the following for
the combinations of coefficients we will need in the calculation,
\begin{eqnarray}
A_{\wavevars,\ell} D_{\wavevars,\ell} &=& 
\frac{1}{\sigma}\frac{\Gamma(\nu(\wavevars)-\ell +1)}
{\Gamma(\nu(\wavevars)+\ell +1)} \cr
\frac{C_{\wavevars,\ell}}{D_{\wavevars,\ell}} &=& -
\frac{
(\sigma^2-1)Q_{\nu(\wavevars)}^\ell{}'\left(\tfrac{1}{\sigma}\right)
K_{\sigma \ell}\left(\wavenumber r_0\right) +
\sigma \kappa r_0 Q_{\nu(\wavevars)}^\ell\left(\tfrac{1}{\sigma}\right)
K_{\sigma \ell}'\left(\wavenumber r_0\right)
}{
(\sigma^2-1)P_{\nu(\wavevars)}^\ell{}'\left(\tfrac{1}{\sigma}\right)
K_{\sigma \ell}\left(\wavenumber r_0\right) +
\sigma \kappa r_0 P_{\nu(\wavevars)}^\ell\left(\tfrac{1}{\sigma}\right)
K_{\sigma \ell}'\left(\wavenumber r_0\right)
}
\cr
B_{\wavevars,\ell} &=& -\frac{
(\sigma^2-1)P_{\nu(\wavevars)}^\ell{}'\left(\tfrac{1}{\sigma}\right)
I_{\sigma \ell}\left(\wavenumber r_0\right) +
\sigma \kappa r_0 P_{\nu(\wavevars)}^\ell\left(\tfrac{1}{\sigma}\right)
I_{\sigma \ell}'\left(\wavenumber r_0\right)
}{
(\sigma^2-1)P_{\nu(\wavevars)}^\ell{}'\left(\tfrac{1}{\sigma}\right)
K_{\sigma \ell}\left(\wavenumber r_0\right) +
\sigma \kappa r_0 P_{\nu(\wavevars)}^\ell\left(\tfrac{1}{\sigma}\right)
K_{\sigma \ell}'\left(\wavenumber r_0\right)
} \,,
\end{eqnarray}
where prime denotes a derivative with respect to the function's argument.

As shown in Refs.\ 
\cite{PhysRevD.59.064017,Khusnutdinov:2004ux,PhysRevD.110.105009},
for $r<r_0$ the wave equation in the string background 
for angular momentum channel $\ell$ can be rewritten 
in terms of the rescaled field $\displaystyle
\varphi_{\wavevars,\ell}(r_*) = \sqrt{r} \psi_{\wavevars,\ell}(r)$
and the physical distance variable
$\displaystyle 
r_* = \frac{r_0}{\sqrt{\sigma^2 - 1}} \arccos \frac{1}{\sigma p(r)}$
as
\begin{equation}
\left(-\frac{d^2}{dr_*^2} + 
\frac{\ell^2-\frac{1}{4}}{r_*^2} + V_\ell(r)
+ \wavenumbersq \right) \varphi_{\wavevars,\ell}(r_*) = 0\,,
\end{equation}
where the potential is
\begin{equation}
V_\ell(r) = \frac{\left(\ell^2-\frac{1}{4}\right) \sigma ^2}{r^2}
-\frac{\left(\ell^2-\frac{1}{4}\right) 
\left(\sigma^2-1\right)}{r_0^2 
\left(\arccos\frac{1}{\sigma p(r)}\right)^2}
+ \frac{(8 \xi - 1) \left(\sigma ^2 - 1\right)}{4 r_0^2} \,.
\end{equation}
For use in the renormalization calculation below, we note that this
potential can be expanded in the curvature,
\begin{equation}
V_\ell(r) = 
\frac{(\ell^2 -1 + 6 \xi) (\sigma^2-1)}{3 r_0^2}
+\frac{r^2 (\ell^2 - \frac{1}{4})(\sigma^2-1)^2}{15 r_0^4}
+ {\cal O}\left[\left(\frac{\sigma^2-1}{r_0^2}\right)^3\right]\,.
\label{eqn:Vexpansion}
\end{equation}
We will make use of this expansion to determine renormalization
counterterms that emerge from analysis of the corresponding
scattering data.

\section{Renormalized Expectation Values and Anomaly}
\label{sec:renorm}

We begin by computing expectation values that make up the
energy density, using results from Refs.\ 
\cite{OLUM2003175,PhysRevD.110.105009},
adapted here to the case of a configuration that is symmetric in $m=2$
dimensions and constant in $n=1$ dimension.  

For $z=z'$, the Green's function can be written as
$G_\sigma(\bm{r},\bm{r}',\kappabar)$,
where $\kappabar = \wavenumber$, with the
corresponding wavefunctions  denoted as $\psi_{\kappabar,\ell}(r) =
\psi_{\wavevars,\ell}(r)$. The expectation values can then be
expressed in terms of this Green's function at coincident points as
\begin{equation}
\left\langle (\partial_t \phi)^2 \right\rangle = 
-\frac{1}{4\pi} \int_\mu^\infty 
\kappabar(\kappabar^2 - \mu^2)
\, d\kappabar  \left[G_\sigma(r,r,\kappabar) - G^{\rm free}(r,r,\kappabar)
+ \frac{{\cal R}}{4\pi\kappabar^2} \left(\xi - \frac{1}{6}\right)
- \frac{{\cal R}^2}{120\pi \kappabar^2} (1 - 10 \xi + 30 \xi^2) f(\kappabar,M)
\right] \,,
\label{eqn:kappaterm}
\end{equation}
where the last three terms represent free space, linear, and quadratic
counterterms, respectively.  As shown in Ref.\
\cite{PhysRevD.110.105009}, the counterterm linear in ${\cal R}$
arises as the limit of the divergence as $m\to 2$ of the free Green's
function at $r\to 0$ times an explicit factor of $m-2$, all of which
multiplies the first-order term in Eq.\ (\ref{eqn:Vexpansion}) for
$\ell=0$.  Because both the second-order contribution to the potential
and the first-order wavefunctions vanish at $r=0$, no such subtlety
arises at second order, and the counterterm is simply given in terms
of the second-order heat kernel coefficient \cite{BrlDv}, which for
our configuration with constant curvature becomes
\begin{equation}
\frac{1}{180} R_{\alpha\beta\gamma\delta} R^{\alpha\beta\gamma\delta}
-\frac{1}{180} R_{\alpha\beta} R^{\alpha\beta}
+ \frac{1}{2}\left(\frac{1}{6} - \xi\right)^2 {\cal R}^2
+\frac{1}{6} \left(\frac{1}{5}-\xi\right) \Box {\cal R}
= \frac{{\cal R}^2}{60}(1 - 10 \xi + 30 \xi^2) \,.
\end{equation}
It is multiplied by the kinematic function arising from second-order
perturbation theory \cite{OLUM2003175}
\begin{equation}
f(\kappabar, M) = \frac{1}{4\kappabar^2 - M^2}
\left(1 + \frac{4\kappabar^2 \arctan\frac{M}{\sqrt{4\kappabar^2 - M^2}}}
{M\sqrt{4\kappabar^2 - M^2}}\right),
\end{equation}
which is written in terms of the renormalization scale $M$.  To avoid
infrared singularities, this scale is typically set equal to the
threshold $2\mu$, or to an imaginary value appropriate to a
relevant scale of the system if $\mu=0$.

Again using the results of Refs.\
\cite{OLUM2003175,PhysRevD.110.105009}, we obtain for the second
derivative term
\begin{equation}
\left\langle \frac{1}{r^2} {\cal D}_r^2 (\phi^2) \right\rangle
= \frac{1}{2\pi} \int_\mu^\infty d\kappabar \, \kappabar \left(
\frac{1}{r^2} {\cal D}_r^2 G_\sigma(r,r, \kappabar) - 
\frac{1}{r^2} {\cal D}_r^2 G^{\rm free}(r,r, \kappabar)
- \frac{1}{4\pi} {\cal R}\right) \,,
\label{eqn:derivterm}
\end{equation}
using the first-order counterterm obtained in Ref.\
\cite{PhysRevD.110.105009} for the $2+1$ dimensional case.  In that
case, the geometry is conformally flat and the combined counterterm
for the energy density from Eqs.\ (\ref{eqn:kappaterm}) and
(\ref{eqn:derivterm}) is given by the $1+1$ dimensional anomaly
contribution $\displaystyle \frac{{\cal R}}{48\pi}$ 
\cite{PhysRevD.13.2720,1977RSPSA.354...59D,PhysRevD.15.2088}.  The
$3+1$ dimensional string is not conformally flat, and correspondingly
Eq.\ (\ref{eqn:derivterm}) has an additional factor of $2$ relative to
Eq.\ (\ref{eqn:kappaterm}), leading to a more complex expression for the
first-order counterterm.  In general we would also
need a wavefunction renormalization counterterm as well, proportional to 
the second derivative of the curvature, but in our case this
counterterm vanishes because because the curvature is constant.
Note that the derivative of the free Green's function at coincident
points vanishes, since it depends only on the difference of its
arguments, but we will find it convenient to include this term for
subsequent manipulations.

Finally, if the field has nonzero mass, we will also need to compute
\begin{equation}
\left\langle \phi^2 \right\rangle
= \frac{1}{2\pi} \int_\mu^\infty d\kappabar \, \kappabar\left(
G_\sigma(r,r, \kappabar) - G^{\rm free}(r,r, \kappabar)\right) \,.
\label{eqn:phisq}
\end{equation}
These expectation values are the ingredients we will need to compute
$\langle T_t^t \rangle = \langle T_z^z \rangle$, along with the sum
of pressures
\begin{eqnarray}
\langle T_r^r + T_\theta^\theta \rangle &=& \left\langle
\frac{1}{r^2}({\cal D}_r \phi)^2 + \frac{1}{r^2}(\partial_\theta \phi)^2
- \left(\frac{1}{2} - \xi \right) \frac{1}{r^2} {\cal D}_r^2 (\phi^2)
+ \xi R \phi^2 \right\rangle \cr
&=& \left\langle (\partial_t \phi)^2 -(\partial_z \phi)^2 + 
\frac{\xi}{r^2} {\cal D}_r^2 (\phi^2) - \mu^2 \phi^2 \right\rangle + 2{\cal A}
= \left\langle 2(\partial_t \phi)^2 +
\frac{\xi}{r^2} {\cal D}_r^2 (\phi^2) - \mu^2 \phi^2 \right\rangle + 2{\cal A}
\,,
\end{eqnarray}
and the trace of the stress-energy tensor
\begin{equation}
\langle T_\alpha^\alpha\rangle = \left\langle 
\left(3\xi-\frac{1}{2}\right)\frac{1}{r^2} {\cal D}_r^2 (\phi^2) - \mu^2 \phi^2
\right\rangle + 4{\cal A}\,.
\end{equation}

Next we turn to the anomalous contribution ${\cal A}$ in 
Eq.\ (\ref{eqn:anomalousrelation}).  It arises because while the
operator expectation values formally agree, the right-hand side
depends on the renormalization scale through the second-order
counterterm contained within the expectation values
$\langle (\partial_t \phi)^2 \rangle$ and 
$\langle (\partial_z \phi)^2 \rangle$, while the left-hand side has no
such dependence.  If we set the renormalization scale to be $2\mu$ and
integrate up to a cutoff $\Lambda$, this contribution becomes
\begin{eqnarray}
{\cal A}_0 &=& \frac{1}{8\pi} \int_\mu^\Lambda d\kappabar \,
\frac{{\cal R}^2}{120 \pi \kappabar^2} (1 - 10 \xi + 30 \xi^2) 
\kappabar(\kappabar^2 - \mu^2) f(\kappabar, 2\mu) \cr
&=& \frac{{\cal R}^2}{3840\pi^2}  (1 - 10 \xi + 30 \xi^2) 
\int_\mu^\Lambda \frac{d\kappabar}{\kappabar}
\left(1 + \frac{\kappabar^2 \arctan\frac{\mu}{\sqrt{\kappabar^2 - \mu^2}}}
{\mu\sqrt{\kappabar^2 - \mu^2}}\right) \cr
&=& \frac{{\cal R}^2}{3840\pi^2} (1 - 10 \xi + 30 \xi^2) 
\left[\frac{\sqrt{\Lambda^2 - \mu^2}}{\mu} 
\arctan \frac{\mu}{\Lambda^2 - \mu^2}
 + \log \left( \frac{\Lambda}{\mu}\right)^2 \right] \,.
\end{eqnarray}
We can recognize the second term in brackets as the logarithmic
divergence being subtracted by renormalization.  In the limit $\Lambda
\gg \mu$, the first term in brackets goes to $1$, leaving the finite
contribution
\begin{equation}
{\cal A}= \frac{{\cal R}^2}{3840\pi^2} (1 - 10 \xi + 30 \xi^2) \,.
\end{equation}
A similar anomalous contribution arises for solitons in $1+1$
dimensional flat spacetime \cite{GRAHAM2024138638}, where the full
calculation can be carried out analytically in dimensional
regularization.

For conformal coupling $\displaystyle \xi=\frac{1}{6}$ and $\mu =
0$, the trace of the stress-energy tensor $\langle T_\alpha^\alpha
\rangle$ becomes equal to $4{\cal A}$.  We therefore obtain the
correct anomaly
\begin{equation}
\langle T_\alpha^\alpha \rangle = 
\frac{1}{2880\pi^2} \left(
R_{\alpha \beta \gamma \delta} R^{\alpha \beta \gamma \delta}
- R_{\alpha \beta} R^{\alpha \beta} - \Box {\cal R} \right)
= \frac{{\cal R}^2}{5760 \pi^2} = 4{\cal A}
\label{eqn:anomaly}
\end{equation}
in that case.  Note that if the curvature within the string were not
constant, the additional counterterm needed in the derivative term
would provide the $\Box {\cal R}$ contribution to the anomaly.

\section{Calculation Using Point String}
\label{sec:point}

While formally correct, the expressions given so far are not yet in a
suitable form for actual calculations, because the mismatch between
the angular momentum quantum numbers $\sigma \ell$ and $\ell$
in the full and free Green's functions respectively
prevents us from taking the difference of the sums term by term.  To
avoid this problem, we add and subtract the contribution of a zero
radius ``point string,'' chosen so that its angular momentum matches
the full Green's function. The mismatched quantum numbers are then
contained in the difference between the point string Green's function
and the free Green's function, which can be efficiently computed via
analytic continuation in $\ell$ \cite{PhysRevD.110.105009}.

The scattering solutions for the point string can be obtained using
the same techniques as for a conducting wedge
\cite{PhysRevD.20.3063,BREVIK1996157}, but with periodic rather than
perfectly reflecting boundary conditions.  The resulting normalized
scattering functions for the point string are
$\psi_{\kappabar,\ell}^{\rm reg,point}(r) = 
I_{\sigma \ell}\left(\kappabar r\right)$ and
$\psi_{\kappabar,\ell}^{\rm out,point}(r) = 
K_{\sigma \ell}\left(\kappabar r\right)$,
and the Green's function becomes
\cite{PhysRevD.34.1918,PhysRevD.35.536,PhysRevD.35.3779}
\begin{equation}
G_\sigma^{\rm point}(\bm{r},\bm{r}',\kappabar) = \frac{\sigma}{\pi}
\sum_{\ell=0}^\infty{}' I_{\sigma \ell}\left(\kappabar r_<\right)
K_{\sigma \ell}\left(\kappabar r_>\right)
\cos \left[\ell \sigma (\theta-\theta')\right] \,.
\label{eqn:pointstring}
\end{equation}
Setting $\sigma=1$, we obtain the free Green's function
\begin{equation}
G^{\rm free}(\bm{r},\bm{r}',\kappabar) = \frac{1}{\pi}
\sum_{\ell=0}^\infty{}' I_{\ell}\left(\kappabar r_<\right)
K_{\ell}\left(\kappabar r_>\right)
\cos \left[\ell (\theta-\theta')\right]
= \frac{1}{2\pi} K_0\left(\kappabar\left|r
e^{i\theta}-r' e^{i\theta'}\right|\right) \,.
\label{eqn:free}
\end{equation}

To calculate the difference between these expressions in the limit where
the points coincide, we can rewrite the $\ell$ sum as a contour
integral \cite{PhysRevD.110.105009}
\begin{equation}
\Delta G_\sigma^{\rm point}(r,r,\kappabar)
= G_\sigma^{\rm point}(r,r,\kappabar) - G^{\rm free}(r,r,\kappabar)
= \frac{1}{\pi^2} 
\int_0^\infty d\lambda K_{i\lambda}\left(\kappabar r\right)
K_{i\lambda}\left(\kappabar r\right)
\frac{\sinh \left[\frac{\lambda}{\sigma}
\pi (\sigma-1)\right]} {\sinh \frac{\lambda}{\sigma}\pi} \,,
\label{eqn:deltapointstring}
\end{equation}
yielding an expression that is well-behaved computationally.
We note that for a massless scalar in three space dimensions, the
complete calculation of the stress-energy tensor for the point string
can be carried out exactly, giving
\cite{PhysRevD.34.1918,PhysRevD.35.536,PhysRevD.35.3779}
\begin{equation}
\langle T_\alpha^\beta \rangle {}^{\rm point,\mu=0} =
\frac{1}{1440 \pi^2 r^4} \left[
(\sigma^4-1) \, {\rm diag}(1,1,-3,1)
+ 20 (6 \xi -1)(\sigma^2 - 1) \, 
{\rm diag}\left(1,-\frac{1}{2},\frac{3}{2},1\right) \right]
\label{eqn:pointanalytic}
\end{equation}
for $\mu=0$, with coordinates listed in the order $t$, $r$,
$\theta$, $z$.

In subtracting and adding back the point string contribution, we can
choose the values of both $\sigma$ and $r$ that we use for the point
string.  While the result of the calculation is independent of this
choice, since we add and subtract the same quantity, it is
advantageous to choose values for which the subtraction is most
effective at improving the numerical convergence.  There are
additional subtleties that arise for the difference of the radial and
angular pressures, which we discuss below.

For all of these calculations, in calculating at radius $r$ in the
ballpoint pen background we will choose $\tilde \sigma = p(r) \sigma$
and $\tilde r = p(r) r$ for the point string.  Note that this radius
differs slightly from the physical radius $r_*$ used in Ref.\
\cite{PhysRevD.110.105009}, which is the integral of $p(r)$ with
respect to $r$ rather than its product with $r$.  The present choice
avoids the need to subtract the additional logarithmic correction
$\displaystyle \frac{1}{2\pi} \log \frac{r_*}{r p(r)}$ discussed
there, which becomes particularly advantageous for the pressure
difference calculation below. Throughout the calculation, we therefore
replace
\begin{equation}
G_\sigma(r,r, \kappabar) - G^{\rm free}(r,r, \kappabar) = 
G_\sigma(r,r, \kappabar) - G_{\tilde \sigma}^{\rm point}
(\tilde r, \tilde r, \kappabar)
+\Delta G_{\tilde \sigma}^{\rm point}(\tilde r, \tilde r, \kappabar)
\label{eqn:gsub}
\end{equation}
and then carry out the subtraction term by term in the Green's
function sum, while using Eq.\ (\ref{eqn:deltapointstring})
to compute the last term in Eq.\ (\ref{eqn:gsub}).  For the difference
of the first two terms we have \cite{PhysRevD.110.105009}
\begin{eqnarray}
G_\sigma(r,r,\kappabar)
-G_{\tilde \sigma}^{\rm point}(\tilde r,\tilde r,\kappabar) &=&
\frac{1}{\pi}\sum_{\ell=0}^\infty{}' \Bigg[
\frac{\Gamma(\nu(\kappabar)-\ell +1)}
{\Gamma(\nu(\kappabar)+\ell +1)}
P_{\nu(\kappabar)}^\ell\left(\tfrac{1}{\sigma  p(r)}\right)
\Bigg(\frac{C_{\kappabar,\ell}}{D_{\kappabar,\ell}}
P_{\nu(\kappabar)}^\ell\left(\tfrac{1}{\sigma  p(r)}\right)
+ Q_{\nu(\kappabar)}^\ell\left(\tfrac{1}{\sigma  p(r)}\right)\Bigg) 
\cr && 
- \tilde \sigma I_{\tilde \sigma \ell}\left(\kappabar \tilde r\right)
K_{\tilde \sigma \ell}\left(\kappabar \tilde r\right)\Bigg]
\end{eqnarray}
for $r<r_0$ and
\begin{equation}
G_\sigma(r,r,\kappabar) -G_\sigma^{\rm point}(r,r,\kappabar)
= \frac{\sigma}{\pi} \sum_{\ell=0}^\infty{}' B_{\kappabar,\ell}
K_{\sigma \ell}\left(\kappabar r\right)
K_{\sigma \ell}\left(\kappabar r\right)
\end{equation}
for $r>r_0$, where both expressions are in the limit of coincident points.

When working with the derivative of this
expression as in Eq.\ (\ref{eqn:derivterm}), we make the replacement
\begin{equation}
\frac{1}{r^2} {\cal D}_r^2 G_\sigma(r,r, \kappabar) - 
\frac{1}{r^2} {\cal D}_r^2 G^{\rm free}(r,r, \kappabar) =
\frac{1}{r^2} {\cal D}_r^2 G_\sigma(r,r, \kappabar) - 
\frac{1}{r^2} \bar{\cal D}_r^2 G_{\tilde \sigma}^{\rm point}
(\tilde r, \tilde r, \kappabar)
+\frac{1}{r^2} \bar{\cal D}_r^2 \Delta G_{\tilde \sigma}^{\rm point}
(\tilde r, \tilde r, \kappabar)
\end{equation}
where $\displaystyle \bar{\cal D}_r = r \frac{d}{dr}$ is the radial
derivative for the point string.  (When acting on the free Green's
function, $\bar{\cal D}_r$ and ${\cal D}_r$ are equivalent
because this derivative is zero.)  Here we have denoted
$\displaystyle \bar{\cal D}_r^2 G_{\tilde \sigma}^{\rm point}
(\tilde r, \tilde r, \kappabar) =
\bar{\cal D}_r^2 G_{\tilde \sigma}^{\rm point}(r, r,
\kappabar)\vert_{r=\tilde r}$, and similarly for other derivatives at
$\tilde r$.
To compute the derivatives of the Green's function, we then use that
for any pair of solutions $\psi^A(r)$ and $\psi^B(r)$ obeying 
Eq.\ (\ref{eqn:eom}), we have
\begin{equation}
\frac{1}{2 r^2} {\cal D}_r^2
\left(\psi_{\kappabar,\ell}^{A} (r)
\psi_{\kappabar,\ell}^{B}(r)\right)
= \left(\frac{(\sigma \ell)^2}{r^2} + \xi {\cal R} + \mu^2
+ \kappabar^2\right)
\psi_{\kappabar,\ell}^{A}(r) \psi_{\kappabar,\ell}^{B}(r)
+ \frac{1}{r^2}\left({\cal D}_r 
\psi_{\kappabar,\ell}^{A}(r)\right)
\left({\cal D}_r\psi_{\kappabar,\ell}^{B}(r)\right) \,,
\end{equation}
and by using recurrence relations we can simplify
\begin{equation}
{\cal D}_r Z_{\nu(\kappabar)}^\ell =
\frac{r\sqrt{\sigma^2-1}}{r_0}
Z_{\nu(\kappabar)}^{\ell+1}\left(\tfrac{1}{\sigma  p(r)}\right)
+ \frac{\ell}{p(r)}
Z_{\nu(\kappabar)}^\ell\left(\tfrac{1}{\sigma  p(r)}\right) \,,
\end{equation}
where $Z$ is either $P$ or $Q$, and similarly for 
derivatives of Bessel functions as well.

We thus obtain the expectation values
\begin{eqnarray}
\left\langle (\partial_t \phi)^2 \right\rangle &=& 
-\frac{1}{4\pi} \int_\mu^\infty 
\kappabar(\kappabar^2 - \mu^2)
\, d\kappabar  \Bigg[
G_\sigma(r,r, \kappabar) - G_{\tilde \sigma}^{\rm point}
(\tilde r, \tilde r, \kappabar)
+\Delta G_{\tilde \sigma}^{\rm point}(\tilde r, \tilde r, \kappabar) 
\cr  &&
+ \frac{{\cal R}}{4\pi\kappabar^2} \left(\xi - \frac{1}{6}\right)
- \frac{{\cal R}^2}{120\pi \kappabar^2} (1 - 10 \xi + 30 \xi^2) f(\kappabar,M)
\Bigg] \cr
\left\langle \frac{1}{r^2} {\cal D}_r^2 (\phi^2) \right\rangle
&=& \frac{1}{2\pi} \int_\mu^\infty d\kappabar \, \kappabar \left(
\frac{1}{r^2} {\cal D}_r^2 G_{\sigma}(r,r, \kappabar)
-\frac{1}{r^2} \bar{\cal D}_r^2 G_{\tilde \sigma}^{\rm point}
(\tilde r, \tilde r, \kappabar)
+\frac{1}{r^2} \bar{\cal D}_r^2 \Delta G_{\tilde \sigma}^{\rm point}
(\tilde r, \tilde r, \kappabar) - \frac{1}{4\pi} {\cal R}\right) \cr
\left\langle \phi^2 \right\rangle
&=& \frac{1}{2\pi} \int_\mu^\infty d\kappabar \, \kappabar\left(
G_\sigma(r,r, \kappabar) - G_{\tilde \sigma}^{\rm point}
(\tilde r, \tilde r, \kappabar)
+\Delta G_{\tilde \sigma}^{\rm point}(\tilde r, \tilde r, \kappabar)\right)
\,.
\end{eqnarray} 
As described above, in all of these expressions the difference of
Green's functions
$G_\sigma(r,r, \kappabar) - G_{\tilde \sigma}^{\rm point}
(\tilde r, \tilde r, \kappabar)$ can be taken term by term in the sum,
while the $\Delta G_{\tilde \sigma}^{\rm point}(\tilde r, \tilde r,
\kappabar)$ contribution is finite on its own, and is given by Eq.\
(\ref{eqn:pointanalytic}) for the case of $\mu=0$.

\section{Pressure Difference}
\label{sec:diff}

The last remaining quantity to compute is the pressure difference
\begin{eqnarray}
\langle T_r^r - T_\theta^\theta \rangle &=& \left\langle
\frac{1}{r^2}({\cal D}_r \phi)^2 - \frac{1}{r^2}(\partial_\theta \phi)^2
- \frac{\xi}{r^2} {\cal D}_r^2 (\phi^2)
+ \frac{2\xi}{r^2} \frac{1}{p(r)} {\cal D}_r (\phi^2) \right\rangle
= -r \partial_r \langle T_r^r\rangle \,,
\label{eqn:pressurediff}
\end{eqnarray}
where the second relation represents the only nontrivial
component of the conservation equation for the stress-energy tensor,
and follows from the equation of motion, Eq.\ (\ref{eqn:eomphi}).
The calculation of this quantity proceeds similarly to those above, but
involves additional subtleties:  First, this calculation is much more
sensitive to the choice of $\tilde \sigma$ and $\tilde r$ values for
the point string.  As a result, to obtain a tractable calculation we
must use precisely the values chosen above, while in the previous
calculations this choice was merely a matter of numerical
optimization.  We also find that we need a scaling correction, similar
to the logarithmic term described above, that depends sensitively on
this choice.

The expectation value $\langle {\cal D}_r^2 (\phi^2) \rangle$
has already been computed above, and the expectation value
$\langle {\cal D}_r (\phi^2) \rangle$ is fully renormalized by
subtracting the free contribution, which is again implemented via
adding and subtracting the point string contribution,
\begin{equation}
\left\langle \frac{2\xi}{r^2} \frac{1}{p(r)}
{\cal D}_r (\phi^2) \right\rangle
= \frac{2\xi}{r^2} \frac{1}{p(r)} \cdot \frac{1}{2\pi} 
\int_\mu^\infty d\kappabar \, \kappabar \left(
{\cal D}_r G_{\sigma}(r,r, \kappabar)
-\bar{\cal D}_r G_{\tilde \sigma}^{\rm point}
(\tilde r, \tilde r, \kappabar)
+\bar{\cal D}_r \Delta G_{\tilde \sigma}^{\rm point}
(\tilde r, \tilde r, \kappabar)\right)\,,
\end{equation}
where the derivative of the Green's function at coincident points
yields a product rule for the two wavefunctions, 
${\cal D}_r \left[\psi_{\kappabar,\ell}^{A} (r)
\psi_{\kappabar,\ell}^{B}(r)\right] =
\left[{\cal D}_r \psi_{\kappabar,\ell}^{A} (r)\right]
\psi_{\kappabar,\ell}^{B}(r) +
\psi_{\kappabar,\ell}^{A} (r)
\left[{\cal D}_r \psi_{\kappabar,\ell}^{B}(r)\right]$.

Finally, the only other expectation value we need to compute is,
for $r<r_0$,
\begin{eqnarray}
\left\langle
\frac{1}{r^2}({\cal D}_r \phi)^2 - \frac{1}{r^2}(\partial_\theta \phi)^2
\right\rangle &=&  
\frac{1}{2\pi r^2} \int_\mu^\infty d\kappabar \kappabar \Bigg(
\frac{1}{\pi}\sum_{\ell=0}^\infty{}' \Bigg\{
\left[ {\cal D}_r
P_{\nu(\kappabar)}^\ell\left(\tfrac{1}{\sigma  p(r)}\right)\right]
\left[ {\cal D}_r R_{\nu(\kappabar)}^\ell\left(\tfrac{1}{\sigma  p(r)}
\right)\right] \\ && 
- (\sigma \ell)^2 P_{\nu(\kappabar)}^\ell\left(\tfrac{1}{\sigma  p(r)}
\right)
R_{\nu(\kappabar)}^\ell\left(\tfrac{1}{\sigma  p(r)}\right)
\cr &&
- \tilde \sigma \left[\kappabar^2 \tilde r^2
I_{\tilde \sigma \ell}'\left(\kappabar \tilde r\right)
K_{\tilde \sigma \ell}'\left(\kappabar \tilde r\right)
+ (\tilde \sigma\ell)^2 
I_{\tilde \sigma \ell}\left(\kappabar \tilde r\right)
K_{\tilde \sigma \ell}\left(\kappabar \tilde r\right)\right]
\Bigg\} + \left(\xi - \frac{1}{6}\right)\frac{{\cal R}}{4\pi}
\cr &&
+ \frac{1}{\pi^2} \int_0^\infty d\lambda 
\left[\kappabar^2 \tilde r^2 
K_{i\lambda}'\left(\kappabar r\right)
K_{i\lambda}'\left(\kappabar \tilde r\right)
+\lambda^2 K_{i\lambda}\left(\kappabar r\right)
K_{i\lambda}\left(\kappabar \tilde r\right)
\right]
\frac{\sinh \left[\frac{\lambda}{\tilde \sigma}
\pi (\tilde\sigma-1)\right]} {\sinh \frac{\lambda}{\tilde\sigma}\pi}
\Bigg) \,, \nonumber
\label{eqn:diffinside}
\end{eqnarray}
where we have defined
\begin{equation}
R_{\nu(\kappabar)}^\ell\left(\tfrac{1}{\sigma  p(r)}\right) =
\frac{\Gamma(\nu(\kappabar)-\ell +1)}
{\Gamma(\nu(\kappabar)+\ell +1)}
\Bigg(\frac{C_{\kappabar,\ell}}{D_{\kappabar,\ell}}
P_{\nu(\kappabar)}^\ell\left(\tfrac{1}{\sigma  p(r)}\right)
+ Q_{\nu(\kappabar)}^\ell\left(\tfrac{1}{\sigma  p(r)}\right)\Bigg) \,,
\end{equation}
and, for $r>r_0$,
\begin{eqnarray}
\left\langle
\frac{1}{r^2}({\cal D}_r \phi)^2 - \frac{1}{r^2}(\partial_\theta \phi)^2
\right\rangle &=&  
\frac{1}{2\pi r^2} \int_\mu^\infty d\kappabar \kappabar \Bigg(
\frac{\sigma}{\pi} \sum_{\ell=0}^\infty{}' B_{\kappabar,\ell}
\left[\kappabar^2 r^2
K_{\sigma \ell}'\left(\kappabar r\right)
K_{\sigma \ell}'\left(\kappabar r\right)
- (\sigma\ell)^2 
K_{\sigma \ell}\left(\kappabar r\right)
K_{\sigma \ell}\left(\kappabar r\right)
\right] \\ &&
+  \frac{1}{\pi^2} \int_0^\infty d\lambda \left[\kappa^2 r^2
K_{i\lambda}'\left(\kappabar r\right)
K_{i\lambda}'\left(\kappabar r\right)
+ \lambda^2 K_{i\lambda}\left(\kappabar r\right)
K_{i\lambda}\left(\kappabar r\right)\right]
\frac{\sinh \left[\frac{\lambda}{\sigma}
\pi (\sigma-1)\right]} {\sinh \frac{\lambda}{\sigma}\pi}
\Bigg)\,. \nonumber
\end{eqnarray}
In all of these expressions, we have subtracted the
point string contribution and then added it back via the integral
representation in Eq.\ (\ref{eqn:deltapointstring}).
The counterterm proportional to ${\cal R}$ in Eq.\ (\ref{eqn:diffinside})
represents a scaling factor, similar to the logarithmic term described
above, corresponding to our choice of $\tilde \sigma$ and $\tilde r$
for the point string subtraction.  As a check of this calculation, we
find numerically that these results obey the conservation law in Eq.\
(\ref{eqn:pressurediff}).

\begin{figure}[htbp]
\includegraphics[width=0.45\linewidth]{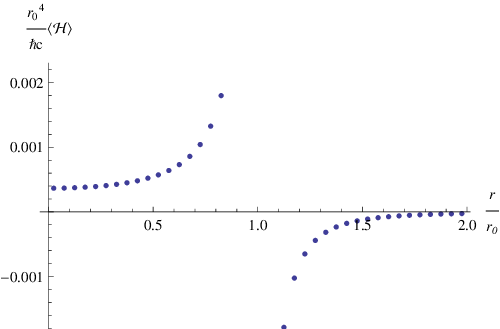} \hfill
\includegraphics[width=0.45\linewidth]{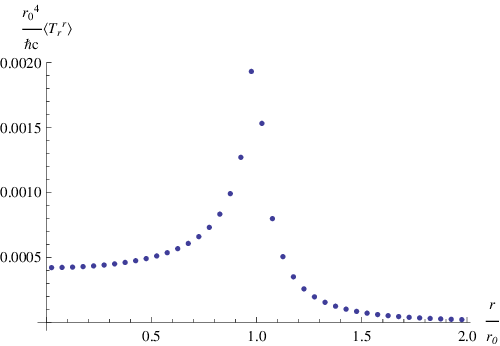}
\includegraphics[width=0.45\linewidth]{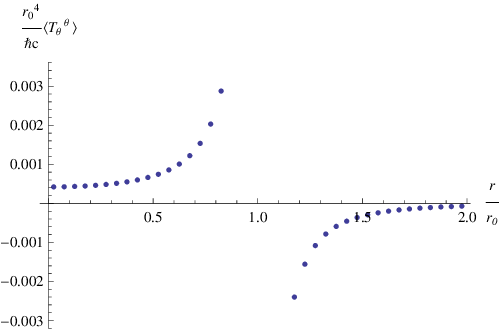} \hfill
\includegraphics[width=0.45\linewidth]{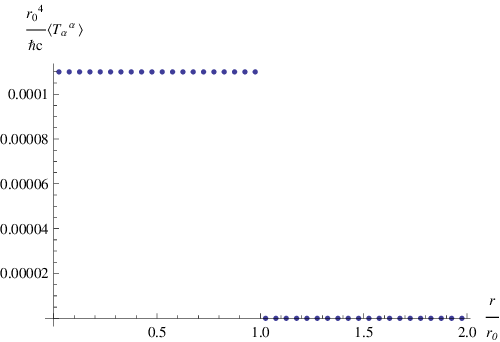}
\caption{Renormalized energy density $\langle \cal H\rangle$, top left,
radial pressure $\langle T_r^r\rangle$, top right, 
angular pressure $\langle T_\theta^\theta\rangle$, bottom left, 
and stress-energy tensor trace $\langle T_\alpha^\alpha\rangle$,
bottom right, all in units of $\displaystyle \frac{\hbar c}{r_0^4}$,
as functions of $r$, in units of $r_0$,
for deficit angle $\displaystyle \theta_0 = \frac{\pi}{3}$,
field mass $\mu=0$, renormalization scale $\displaystyle M = \frac{i}{r_0}$,
and conformal coupling $\displaystyle \xi = \frac{1}{6}$.  The
stress-energy tensor trace arises entirely from the anomaly
contribution, Eq.\ (\ref{eqn:anomaly}).}
\label{fig:conformal}
\end{figure}

\begin{figure}[htbp]
\includegraphics[width=0.45\linewidth]{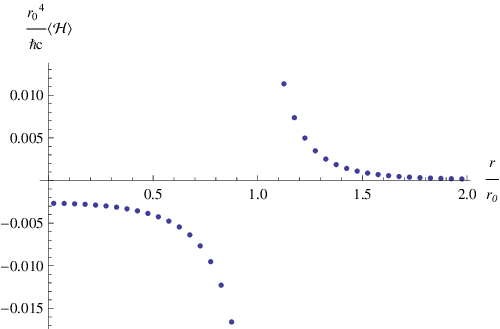} \hfill
\includegraphics[width=0.45\linewidth]{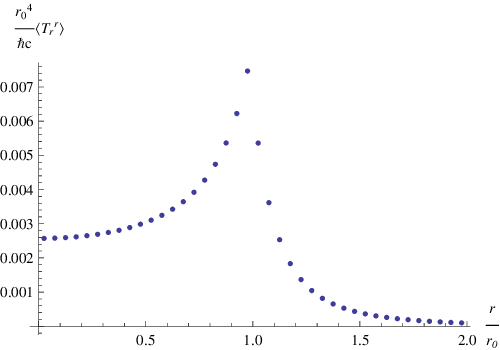}
\includegraphics[width=0.45\linewidth]{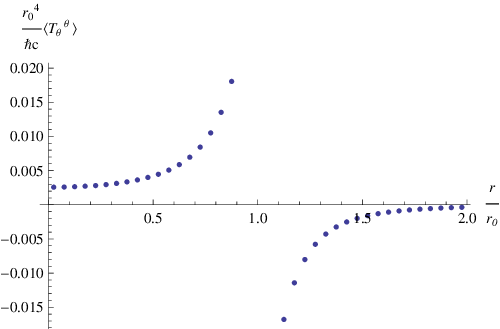} \hfill
\includegraphics[width=0.45\linewidth]{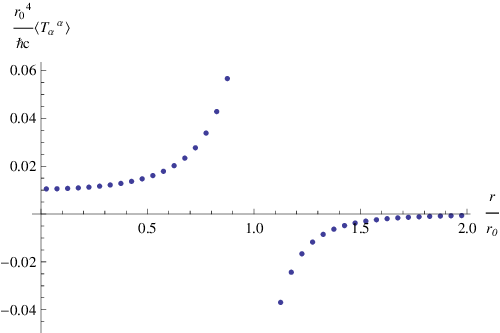}
\caption{Renormalized energy density $\langle \cal H\rangle$, top left,
radial pressure $\langle T_r^r\rangle$, top right, 
angular pressure $\langle T_\theta^\theta\rangle$, bottom left, 
and stress-energy tensor trace $\langle T_\alpha^\alpha\rangle$,
bottom right, all in units of $\displaystyle \frac{\hbar c}{r_0^4}$,
as functions of $r$, in units of $r_0$, 
for $\displaystyle \theta_0 = \frac{\pi}{3}$,
field mass $\mu=0$, renormalization scale $\displaystyle M = \frac{i}{r_0}$,
and minimal coupling $\xi = 0$.}
\label{fig:minimal}
\end{figure}

\begin{figure}[htbp]
\includegraphics[width=0.45\linewidth]{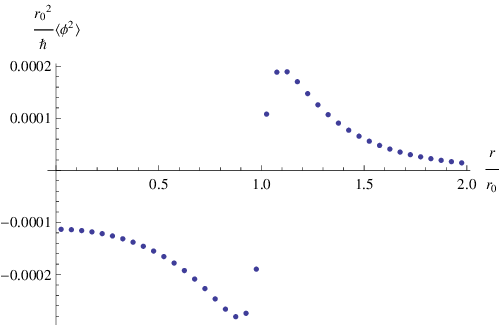} \hfill
\includegraphics[width=0.45\linewidth]{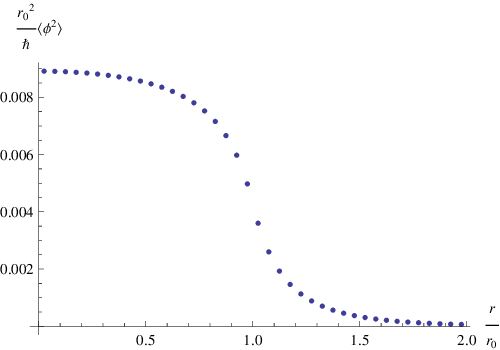}
\caption{Renormalized expectation value $\langle \phi^2 \rangle$,
in units of $\displaystyle \frac{\hbar}{r_0^2}$,
as a function of $r$, in units of $r_0$, 
for $\displaystyle \theta_0 = \frac{\pi}{3}$ and field mass $\mu=0$.
The left panel shows conformal coupling $\displaystyle \xi =
\frac{1}{6}$, while the right panel shows minimal coupling $\xi = 0$.}
\label{fig:phisq}
\end{figure}

\begin{figure}[htbp]
\includegraphics[width=0.45\linewidth]{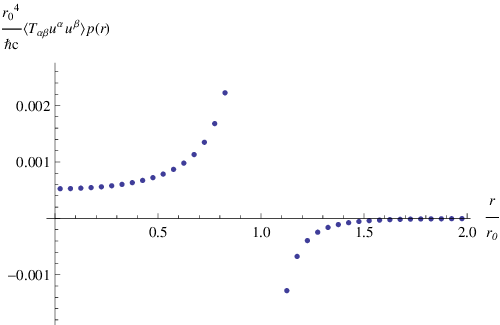} \hfill
\includegraphics[width=0.45\linewidth]{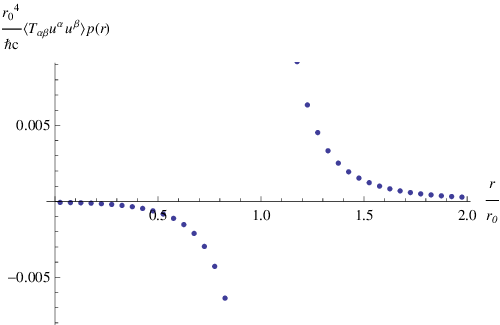}
\caption{Contribution to the averaged null energy condition integral
for a radial geodesic, $\langle T_{\alpha\beta}\rangle u^{\alpha}
u^{\beta} p(r) = \langle {\cal H} +  T_r^r \rangle p(r)$,
in units of $\displaystyle \frac{\hbar c}{r_0^4}$,
as a function of $r$, in units of $r_0$,
for $\displaystyle \theta_0 = \frac{\pi}{3}$, field mass $\mu=0$, and 
renormalization scale $\displaystyle M = \frac{i}{r_0}$.
The left panel shows conformal coupling $\displaystyle \xi =
\frac{1}{6}$, while the right panel shows minimal coupling $\xi = 0$.}
\label{fig:nec}
\end{figure}

\section{Results}
\label{sec:results}

We can now compute renormalized expectation values numerically by
summing over $\ell$ and integrating over $\kappabar$.
Figures \ref{fig:conformal} and \ref{fig:minimal} show the components
of the stress-energy tensor and its trace for a massless
field renormalized at the scale of the string radius, $\displaystyle 
M = \frac{i}{r_0}$, in the case of conformal and minimal coupling
respectively.  For conformal coupling, the trace is given entirely by
the anomaly contribution, Eq.\ (\ref{eqn:anomaly}).  Consistent with
the results from the point string, the radial pressure is always
positive, while the other components switch sign both inside and
outside the string and between conformal and minimal coupling.  The
singularity at $r=r_0$ is an artifact of the discontinuity in the
curvature, which in an actual string would decrease to zero
continuously.  The integrals of these quantities over space, taken as
principal values, remain finite \cite{OLUM2003175}, although they are
difficult to compute numerically because of this behavior.
Numerically, we find that the elements of the stress-energy tensor all
diverge as $\displaystyle \frac{{\cal R}}{(r-r_0)^2}$ for $r\to r_0$,
except for $\langle T_r^r\rangle$, which diverges proportionally to
$\displaystyle \frac{{\cal R}}{r|r-r_0|}$, consistent with
Eq.\ (\ref{eqn:pressurediff}).
In contrast, the expectation value $\langle \phi^2 \rangle$ in 
Eq.\ (\ref{eqn:phisq}) is smooth at $r=r_0$, as shown in Fig.\
\ref{fig:phisq}.  For all of these cases, the behavior at the
boundary matches up analogously to that of a scalar field interacting
with a square well potential in flat spacetime, with the degree of the
power law divergence determined by the codimension of the singular surface
\cite{OLUM2003175}. 

Another quantity of interest is the contribution to the null energy
condition (NEC).  Energy conditions represent restrictions on the
stress-energy tensor that can potentially rule out exotic phenomena,
such as closed timelike curves or superluminal travel.  The null energy
condition, $T_{\alpha \beta} u^{\alpha} u^{\beta} \geq 0$ for
a null 4-velocity $u^{\alpha}$ has known violations, but the weaker
condition in which it is averaged over an achronal null geodesic 
(AANEC) may still be viable, and is sufficient to rule out exotic
phenomena
\cite{Graham:2007va,PhysRevD.81.024038}; while examples that violate
this condition in have been found
\cite{PhysRevD.81.024039,Kontou_2020}, they are expected 
only to exist for quantum fields in curved spacetime
\cite{PhysRevD.54.6233,Bousso:2015wca,Hartman:2016lgu,PhysRevD.54.6233,Kontou:2012ve,Kontou:2015yha}, and it is not
known whether such violations can be self-consistently constructed
from solutions to the Einstein equation.

Here we consider the contribution from quantum fluctuations, to which
we would then add the classical contribution of the matter that is creating
the string.  By Einstein's equation for the classical string
background as given in Eq.\ (\ref{eqn:curvature}), 
the latter has $\displaystyle {\cal H} = -T_{zz} = 
\frac {\cal R}{16 \pi G} > 0$, with other components zero
\cite{1985ApJ...288..422G}, and therefore gives a contribution that
should only reduce the possibility that AANEC is violated.  We
consider a radial geodesic with 4-velocity
$\displaystyle
u^{\alpha} = \begin{pmatrix} 1 & \frac{1}{p(r)} & 0 & 0 \end{pmatrix}$,
which obeys the geodesic equation $u^\alpha \nabla_\alpha u^\beta = 0$.
The integrand of the AANEC integral
\begin{equation}
\int_0^\infty dr \, p(r) \langle T_{\alpha\beta} u^{\alpha} u^{\beta}\rangle =
\int_0^\infty dr \, p(r) \langle {\cal H} + T_r^r\rangle
\end{equation}
is shown in Fig.\ \ref{fig:nec} for
both minimal and conformal coupling.  For minimal coupling, the
integral over the geodesic is difficult to determine because it
appears to be close to zero, with the singularity at $r=r_0$
preventing a detailed calculation, while for conformal coupling the
integral appears to be positive, indicating that the condition is
obeyed by the quantum contribution.

\section{Conclusions}
\label{sec:conclusions}

We have demonstrated a complete calculation of the renormalized quantum
stress-energy tensor in the background of a ``ballpoint pen'' cosmic
string geometry in $3+1$ dimensions.  This example represents one of
the few cases in which quantum effects in curved spacetime can be
calculated in detail, and concretely demonstrates the role played by
the conformal trace anomaly in the renormalization process, for both
the full four-dimensional spacetime in which the calculation is
carried out and the two-dimensional subspace in which it has
nontrivial curvature.  This result also allows one to calculate the
quantum field's contribution to the averaged null energy condition,
which appears to be obeyed.

While these calculations have been carried out in three space
dimensions, it is straightforward to reduce to the previously
considered case of two space dimensions
\cite{PhysRevD.110.105009}: in computing the expectation values
we divide by $\sqrt{\kappabar^2-\mu^2}$ in all of the $\kappabar$
integrands, the $\langle (\partial_t \phi)^2 \rangle$ integral is
multiplied by $4$ while the other integrals are multiplied by $2$,
\cite{OLUM2003175} and all $k_z$, $\partial_z \phi$, second-order
counterterm, and anomaly contributions are set to zero.  (Accordingly,
the coefficient $\displaystyle 3 \xi -\frac{1}{2}$ in the trace of the
stress-energy tensor would become $\displaystyle 2 \xi -\frac{1}{4}$,
vanishing for conformal coupling $\displaystyle \xi = \frac{1}{8}$.)  The
powers of the divergences as $r\to r_0$ would all be reduced by
one as well, with a power of zero corresponding to a log divergence.
It is also straightforward to extend these calculations
to the ``flowerpot'' model, in which the curvature is entirely
concentrated in a $\delta$-function contribution at $r=r_0$, simply by
making the appropriate substitution of scattering wavefunctions
\cite{PhysRevD.42.2669,PhysRevD.110.105009}.  In that case,
however, the curvature is always zero for $r\neq r_0$, so all
counterterms and anomaly contributions vanish, the divergences at
$r=r_0$ do not necessarily cancel as principal values because the
integrated quantities
can be divergent in a singular background, and we cannot analyze
the AANEC for a radial geodesic because it must pass through the
singularity at $r=r_0$.  One could also consider a curvature
background that goes to zero continuously for increasing $r$, rather
than as a step or $\delta$-function.  In that case, the scattering
wavefunctions would likely need to be computed numerically, for
example by using variable phase techniques \cite{density}, and there
would be additional contributions to the second-order counterterm and
corresponding anomaly contribution proportional to $\Box {\cal R}$.

It would also be interesting to calculate the integrals over space of
these local densities.  In flat spacetime, these calculations can
be implemented effectively by making use of the relationship between
the integral over space of the difference between the free and
interacting Green's functions and the Jost function $F_\ell(\kappabar)$
in scattering channel $\ell$ \cite{density},
\begin{equation}
2 \kappa \int_0^\infty 2 \pi r dr\,\left[G_\ell(r,r,\kappabar) - 
G^{(0)}_\ell(r,r,\kappabar)\right] 
= \frac{d}{d\kappabar} \log F_\ell(\kappabar)\,,
\end{equation}
and both exact and approximate expressions for the
ballpoint pen Jost function have been obtained in
Refs. \cite{PhysRevD.59.064017,Khusnutdinov:2004ux}. A similar
approach here would need to take into account the different topology
between the full and free backgrounds.  Such an approach
can potentially avoid the difficulties associated with integrating
over the singularity at $r=r_0$ and provide generally applicable
insights into whether the AANEC is obeyed.

\acknowledgments
It is a pleasure to thank K.\ Olum for sharing preliminary work on
this topic and M.\ Koike, X.\ Laquidain, and H.\ Weigel
for helpful conversations and feedback. 
N.\ G.\ was supported in part by the National
Science Foundation (NSF) through grant PHY-2209582.

\bibliographystyle{apsrev}
\bibliography{flowerpen}

\end{document}